\documentclass{rmaa}

\suppressfulladdresses 

\SetYear{2009}

\title{A perturbative analysis of Quasi-Radial density waves in galactic disks.} 

\author{X. Hernandez\altaffilmark{1} and I. Puerari\altaffilmark{2}}

\altaffiltext{1}{Instituto de Astronom\'\i{}a,
  Universidad Nacional Aut\'onoma de M\'exico, M\'exico (xavier@astroscu.unam.mx).}
\altaffiltext{2}{Instituto Nacional de Astrof\'{\i}sica, Optica y Electr\'onica,
 Santa Mar\'{\i}a Tonantzintla, M\'exico (puerari@inaoep.mx).}

\shortauthor{X. Hernandez and I. Puerari}
\shorttitle{Quasi-Radial density waves in galaxies}

\listofauthors{X. Hernandez}
\indexauthor{X. Hernandez.}

\abstract{ 
The theoretical understanding of density waves in disk galaxies starts from the classical
WKB perturbative analysis of tight-winding perturbations, the key assumption being that
the potential due to the density wave is approximately radial. The above has served as a
valuable guide in aiding the understanding of both simulated and observed galaxies, in spite
of a number of caveats being present. The observed spiral or bar patterns in real galaxies
are frequently only marginally consistent with the tight-winding assumption, often in fact,
outright inconsistent. Here we derive
a complementary formulation to the problem, by treating quasi-radial density waves under
simplified assumptions in the linear regime. We 
assume that the potential due to the density wave is approximately tangential, and derive
the corresponding dispersion relation of the problem. We obtain an instability criterion for the onset of
quasi-radial density waves, which allows a clear understanding of the increased stability of the higher order 
modes, which appear at progressively larger radii, as often seen in real galaxies. The theory
naturally yields a range of pattern speeds for these arms which appears constrained by the condition
$\Omega_{p}<\Omega_{0} \pm \kappa /m$. For the central regions of galaxies where solid body rotation
curves might apply, we find weak bars in the oscillatory regime with various pattern speeds, including
counter rotating ones, and a prediction for $\Omega_{p}$ to increase towards the centre, as seen in the rapidly rotating bars 
within bars of some numerical simulations. We complement this study with detailed numerical simulations
of galactic disks and careful Fourier analysis of the emergent perturbations, which support the theory 
presented.}

\resumen{}

\addkeyword{instabilities}
\addkeyword{stellar dynamics}
\addkeyword{galaxies: kinematics and dynamics}
\addkeyword{galaxies: structure}
\addkeyword{galaxies: spiral}

\begin{document}
\maketitle

\section{Introduction}
\label{sec:intro}

The development of short-wavelength, tight-winding disk wave dynamics, notably by
Lin \& Shu (1964), Shu (1970) and Toomre (1969), still
forms the basic analytical substrate used to understand galactic structure observations 
and numerical simulations of galactic dynamics. This WKB perturbative analysis
exploits the fact that for tight-winding spirals, the long range character of the
gravitational potential due to the density wave one is modeling disappears, the response becomes
local, and an analytical formulation to the problem becomes feasible. The dispersion relation
that results has been used successfully over the decades to gain some understanding of the expected 
physical scalings between pattern speed, the orbital frequency, the epicycle frequency and the
mass density of galactic disks. 

However, several weak points in the application of the classical WKB density wave dynamics
to real galaxies have been recognised since they where originally proposed. First, one would
like spiral arms to be very tightly wound, making the approximation on the radial dominance 
of the perturbed potential clearly valid. For typical early type spirals, spiral arms are clearly 
tightly wound, but often one finds late type spirals with almost radial 'spiral' arms.
The spiral patterns of many galaxies appear to be on the limit of 
applicability for this approximation, and many, galactic bars included, are outright out of it. 

One of the strong points of the density wave theory was the substantial easing of the 
winding problem, which can actually be used to rule out any material arms interpretation
for spiral structure in galaxies, provided one seeks a long lived structure. 

Here we explore a complementary, and
in a certain sense, perpendicular approach. We shall assume that the density wave being treated is
quasi-radial, and that it represents only a small disturbance on the overall potential. Under this
assumptions, the potential of the disturbance will only be felt near the density enhancement, 
and will be tangential, directed towards the density enhancement itself, particularly
if one assumes no strong radial gradients in the amplitude of the density wave itself. 
Treating only the circular motions of a collisionless component, and ignoring the vertical 
gradients of all quantities involved, 
lands us in the thin disk approximation for stellar disks, a scenario often used in galactic dynamical 
works and in studies of galactic density 
waves in analytical developments, orbital structure analysis or N-body simulations, being
Hernandez \& Cervantes-Sodi (2006), Jalali \& Hunter (2005), Maciejewski \& Athanassoula (2008) 
and O'Neil \& Dubinski (2003) examples of the above.

This allows an approximate analytical treatment of the problem which we present here in section (2).
A dispersion relation for the problem appears, the consequences of which we explore under two
idealised galactic regimes, flat rotation curve, and solid body rotation, with the aim of 
sampling the resulting physics under the approximate dynamical conditions where one
finds spiral arms and bars. As in the case of the Lin-Shu formalism, we begin by treating only
a stellar disk, thinking of modeling the old stellar population of a present day galaxy.
The development is idealised and remains within the linear regime, for quasi-radial density waves,
for a disk made up of close to circular orbits. While lacking the detail of more advanced studies
(e.g. Evans \& Read 1998,  Polyachenko 2004, Jalali \& Hunter 2005), it has the advantage of 
identifying clearly a physical instability criterion which captures some of the details missed by 
the Lin-Shu formalism, highlighting the role of stellar diffusion. As an independent confirmation of
the ideas presented, we also run an extensive grid of numerical models, and analyse the details of the
small amplitude density perturbations which appear. It is encouraging that the results of these direct 
numerical simulations validate the general results of the ideas developed in this study.

In section (3), for the case of the flat rotation curve regime, we derive a dispersion relation for
the problem, and a corresponding stability criterion for the onset of quasi-radial density
perturbations, which is seen to be clearly satisfied for any reasonable galactic disk, for
low values of $m$ and $Q$, the symmetry number for the pattern and Toomre's parameter. 
The case of a solid body rotation regime is also treated, where we show that for weak bars
in the oscillatory regime of the dispersion relation, relatively constant values for
$\Omega_{p}$ appear quite naturally, at values which represent 'slow', 'fast' or even counter-rotating
bars, for natural values of $m$ and $Q$. The extent of these bars being naturally 
limited by the corotation condition. A detailed numerical model is presented in section (4), including
a careful Fourier analysis of the density perturbations which appear, yielding results in consistency
with the theory developed. Section (5) presents our conclusions.

\section{Physical Setup}
\label{sec:framework}

In this section we develop the simplified physical model for quasi-radial density waves.
As opposed to the Lin-Shu formalism, designed to model tight-winding waves, we assume that the derivatives of the 
potential due to the perturbation are dominated by the angular component, 
rather than the radial one. We shall start from the angular moments equation written in cylindrical
coordinates $(R, \theta, z)$, in a reference frame which rotates with a constant angular frequency 
$\Omega_{P}$ parallel to the z axis, but without assuming azimuthal symmetry.
In spite of the multi-component nature of a galactic system, we shall
begin by ignoring the gaseous component, in an attempt to model the underlying
density perturbation represented by loosely-wound arms in the old stellar
population. We shall also assume no vertical or radial velocities, in modelling
weak radial arms which only result in weak angular distortions in both the density and
velocity fields. Taking also an isotropic velocity ellipsoid with no vertex deviation, i.e.
$\sigma_{ij} =\delta_{ij}\sigma$, results in the following equation (e.g. Vorobyov \& Theis 2006):

\begin{equation}
\frac{\partial(R \Sigma v_{\theta})}{\partial t} + \frac{1}{R}\frac{\partial(R \Sigma v_{\theta}^{2})}{\partial \theta}
+\frac{\partial (\Sigma \sigma^{2})}{\partial \theta} =-\Sigma \frac{\partial \Phi_{eff}}{\partial \theta}
\end{equation}

In the above equation $\Phi_{ef}=\Phi-|\vec{\Omega_{p}} \times \vec{r}|^{2}/2$
and $\Sigma$ is the mass surface density of the disk. Taking the quasi-radial arms
we are interested in modelling as a perturbation on the axisymmetric centrifugal equilibrium solution
gives rise to the following set of conditions at every radius:

$$
\Sigma = \Sigma_{0} + \epsilon \Sigma_{1}(\theta)   
$$

$$
v_{\theta} = v_{0}(R) + \epsilon v_{1}(\theta)
$$

$$
\sigma  = \sigma_{0} + \epsilon \sigma_{1}(\theta)   
$$

$$
\Phi(R,\theta) = \Phi_{0}(R) + \epsilon \Phi_{1}(R, \theta)
$$

In the above $\Phi_{0}$ is the axially symmetric 
gravitational potential, which cancels the centrifugal force fixing $\vec{V(R)} =(0, R(\Omega_{0}-\Omega_{p}), 0)$, with $\Omega_{0}(R)$ 
the centrifugal equilibrium orbital frequency.  $\Phi_{1}$ will hence be the potential perturbation due to the quasi-radial arms, resulting 
in angular perturbations $\Sigma_{1}(\theta)$, $v_{1}(\theta)$ and $\sigma_{1}(\theta.)$ The form taken for the velocity implies no further
inertial terms are necessary. The unperturbed state trivially satisfies eq.(1) for the standard axisymmetric centrifugal equilibrium
state.

To first order in the perturbation, eq.(1) now becomes:

\begin{equation}
\begin{array}{lll}
 R \Sigma_{0} \frac{\partial v_{1}}{\partial t} + R^{2} \Delta \Omega \frac{\partial \Sigma_{1}}{\partial t}
+2R \Delta \Omega \Sigma_{0} \frac{\partial v_{1}}{\partial  \theta} & & \\
+ (R^{2} \Delta^{2} \Omega +\sigma_{0}^{2})\frac{\partial \Sigma_{1}}{\partial \theta} + 2 \sigma_{0} \Sigma_{0} \frac{\partial \sigma_{1}}{\partial \theta} & &\\
= -\Sigma_{0} \frac{\partial \Phi_{1}}{\partial \theta} & &
\end{array}
\end{equation}

In the above equation we have introduced $\Delta \Omega = (\Omega_{0}-\Omega_{p})$, and
have omitted making explicit the radial dependence of all the above
variables, which remains implicit.

We now introduce an isothermal equation of state (e.g. Binney \& Tremaine 1987) or equivalently the conservation of
phase-space density $\Sigma/\sigma^{3} =cte.$ to eliminate $\sigma_{1}$ for $\Sigma_{1}$ through:

\begin{equation}
\sigma_{1} = \left( \frac{\sigma_{0}}{3 \Sigma_{0}} \right) \Sigma_{1},
\end{equation}

\noindent with which eq.(2) becomes:

\begin{equation}
\begin{array}{lll}
 R \Sigma_{0} \frac{\partial v_{1}}{\partial t} + R^{2} \Delta \Omega \frac{\partial \Sigma_{1}}{\partial t}
+2R \Delta \Omega \Sigma_{0} \frac{\partial v_{1}}{\partial  \theta} & & \\
+ \left( R^{2} \Delta^{2} \Omega + \frac{5}{3}\sigma_{0}^{2} \right) \frac{\partial \Sigma_{1}}{\partial \theta} = 
-\Sigma_{0} \frac{\partial \Phi_{1}}{\partial \theta} & &
\end{array}
\end{equation}

We see the assumption of eq.(3) in the numerical constant of the fifth term in the above equation. Any other similar equation of
state would only change this constant slightly, leaving all conclusions qualitatively unchanged, and only modified to within a small
numerical factor of order unity. The partial angular derivative of the above equation reads:

\begin{equation}
\begin{array}{lll}
 R \Sigma_{0} \frac{\partial^{2} v_{1}}{\partial \theta \partial t} + R^{2} \Delta \Omega \frac{\partial^{2} \Sigma_{1}}{\partial \theta \partial t}
+2R \Delta \Omega \Sigma_{0} \frac{\partial^{2} v_{1}}{\partial  \theta^{2}} & & \\
+ \left( R^{2} \Delta^{2} \Omega + \frac{5}{3}\sigma_{0}^{2} \right) \frac{\partial^{2} \Sigma_{1}}{\partial \theta^{2}} = 
-\Sigma_{0} \frac{\partial^{2} \Phi_{1}}{\partial \theta^{2}} & &
\end{array}
\end{equation}

We will change the dependencies of the above equation on $v_{1}$ for dependencies on $\Sigma_{1}$
through the use of the continuity equation, which reads:

\begin{equation}
\frac{\partial \Sigma}{\partial t} +\frac{1}{R}\frac{\partial(R \Sigma v_{R})}{\partial R} + \frac{1}{R} \frac{\partial (\Sigma v_{\theta})}{\partial \theta}=0.
\end{equation}

Introducing the perturbation conditions with only tangential velocities into the above continuity equation yields:

\begin{equation}
\frac{\partial \Sigma_{1}}{\partial t}+ \frac{\Sigma_{0}}{R} \frac{\partial v_{1}}{\partial \theta} + \Delta \Omega \frac{\partial \Sigma_{1}}{\partial \theta}=0.
\end{equation}

From this last equation we can solve for $\partial v_{1} /\partial \theta$ to construct:

\begin{equation}
\frac{\partial^{2} v_{1}}{\partial \theta^{2}}=- \frac{R \Delta \Omega}{\Sigma_{0}} \frac{\partial^{2}\Sigma_{1}}{\partial \theta^{2}}
- \frac{R}{\Sigma_{0}} \frac{\partial^{2} \Sigma_{1}}{\partial \theta \partial t}
\end{equation}

\begin{equation}
\frac{\partial^{2} v_{1}}{\partial \theta \partial t}=- \frac{R \Delta \Omega}{\Sigma_{0}} \frac{\partial^{2}\Sigma_{1}}{\partial \theta \partial t}
- \frac{R}{\Sigma_{0}} \frac{\partial^{2} \Sigma_{1}}{\partial t^{2}}
\end{equation}

The last two above are now used to replace the dependence on $v_{1}$ appearing in eq.(5) for a dependence on $\Sigma_{1}$, giving:

\begin{equation}
\begin{array}{lll}
- R^{2} \frac{\partial^{2} \Sigma_{1}}{\partial t^{2}} -2 R^{2} \Delta \Omega \frac{\partial^{2} \Sigma_{1}}{\partial \theta \partial t}
+ \left(\frac{5}{3}\sigma_{0}^{2} - R^{2}\Delta^{2} \Omega  \right) \frac{\partial^{2} \Sigma_{1}}{\partial \theta^{2}} & & \\
= -\Sigma_{0} \frac{\partial^{2} \Phi_{1}}{\partial \theta^{2}} & &
\end{array}
\end{equation}

Finally, we relate $\Sigma_{1}$ and $\Phi_{1}$ through the
first order Poisson equation of the problem, which under the assumption of
tangential forces due to the almost radial perturbation dominating over the 
other two directions gives:

\begin{equation}
\frac{1}{R^{2}}\frac{\partial^{2}\Phi_{1} }{\partial \theta^{2}}= 4 \pi G \rho_{1},
\end{equation}

allows to write  eq.(10) as:

\begin{equation}
\begin{array}{lll}
- R^{2} \frac{\partial^{2} \Sigma_{1}}{\partial t^{2}} -2 R^{2} \Delta \Omega \frac{\partial^{2} \Sigma_{1}}{\partial \theta \partial t}
+ \left(\frac{5}{3}\sigma_{0}^{2} - R^{2}\Delta^{2} \Omega  \right) \frac{\partial^{2} \Sigma_{1}}{\partial \theta^{2}} & & \\
= -4 \pi G \rho_{0} \Sigma_{1} & &
\end{array}
\end{equation}

In the above equation we have used $h$, the height scale of the disk, as $h=\Sigma_{0}/\rho_{0} = \Sigma_{1}/\rho_{1}$.

We now turn to well known galactic structure scalings to re-write the right
hand side of eq.(12) in terms of more helpful dynamical variables, specifically, we 
shall make use of vertical virial equilibrium in the disk and Toomre's stability criterion, 

\begin{equation}
\frac{\kappa \sigma_{0}}{\pi G \Sigma_{0}} =Q.
\end{equation}

In the above $\kappa^{2}=4\Omega_{0}^{2}+2\Omega_{0} R (d\Omega_{0}/dR)$ is the epicycle frequency 
as functions of the radius. The star formation processes in the disks of spiral galaxies 
have often been thought of as constituting a self regulated cycle. 
The gas in regions where $Q<$ a certain critical value finds itself in a regime where local 
self-gravity, appearing through $(G \Sigma_{0})$ in eq.(13), dominates over the combined
effects of rotational shears (or equivalently global tidal forces) and the stabilising 
thermal 'pressure' effects of the product $(\kappa \sigma)$. This leads to collapse and the triggering of
star formation processes, which in turn result in significantly energetic events. The above include
radiative heating, the propagation of ionisation fronts, shock waves and in general an efficient
turbulent heating of the gas media, raising $\sigma$ locally to values resulting in $Q>$ a certain critical
threshold. This
restores the gravitational stability of the disk and shuts off the star formation processes. On timescales
longer than the few $\times 10$ million years of massive stellar lifetimes, an equilibrium 
is expected where star formation proceeds at a rate equal to that of gas turbulent dissipation, at
time averaged values of $Q\sim Q_{crit}$. Examples of the above can be found in Dopita \& Ryder (1994), 
Koeppen et al. (1995), Firmani et al. (1996) and Silk (2001).
 
The preceding argument applies to the gaseous component, not 
the stellar one, which is the one we are interested in modeling, particularly the old stellar component.
However, if stars retain essentially the velocity dispersion values of the gas from which they formed,
one also expects $Q \sim Q_{crit}$ for the stars. Alternatively, if we consider the dynamical
heating of the stellar populations in the disk through encounters with giant molecular clouds 
or the spiral arms themselves, slightly larger values of $Q$ would be expected for the stars than for
the gas. We shall therefore take eq.(13)
as valid throughout the disk (e.g. O'Neill \& Dubinski 2003, Maciejewski \& Athanassoula 2008), 
without specifying any particular values of $Q$ at this point. 
Imposing viral equilibrium in the vertical direction in the disk (e.g. Binney \& Tremaine 1987) 
yields:

\begin{equation}
h=\frac{\sigma_{0}^{2}}{2 \pi G \Sigma_{0}},
\end{equation}

Since $h=\Sigma_{0}/\rho_{0}$, we can replace the dependence on $\Sigma_{0}$ for one on $\rho_{0}$ in eqs.(13)
and (14), and solve for $(\sigma_{0}/h)$ in both eqs.(13) and (14) to obtain:

\begin{equation}
\pi G \rho_{0} = 2 \left(\frac{\kappa}{Q} \right)^{2}.
\end{equation}

Equation (15) serves to re-write the right hand side of eq.(12)
and obtain:

\begin{equation}
\begin{array}{lll}
- R^{2} \frac{\partial^{2} \Sigma_{1}}{\partial t^{2}} -2 R^{2} \Delta \Omega \frac{\partial^{2} \Sigma_{1}}{\partial \theta \partial t}
+ \left(\frac{5}{3}\sigma_{0}^{2} - R^{2}\Delta^{2} \Omega  \right) \frac{\partial^{2} \Sigma_{1}}{\partial \theta^{2}} & & \\
= -8 \Sigma_{1} \left( \frac{\kappa}{Q}  \right)^{2}& &
\end{array}
\end{equation}

Equation (16) is now a second order partial differential equation for the temporal and angular variations of the density of the perturbation. 
This is valid at each radius, and shows dependence on both the gravitational dynamics of the system through $\kappa$ and $\Omega$, and on the 
local stability and structure of the disk through $Q$ and $\sigma_{0}$. Now we impose periodic solutions in $\theta$ for the perturbation in density,
with an oscillatory or growing time dependence:

\begin{equation}
\Sigma_{1} \propto e^{-i (m \theta-\omega t)},
\end{equation}

\noindent with $m$ an integer, which when introduced into eq.(16) gives:

\begin{equation}
(\omega -m \Delta \Omega)^{2}= \frac{5}{3} \left( \frac{m \sigma_{0}}{R}\right)^{2} - 2\left( \frac{2 \kappa}{Q}\right)^{2}.
\end{equation}

Equation (18) is the dispersion relation for the problem. For a density wave of the type being described to develop, we require for 
$\omega$ to be imaginary, i.e.,

\begin{equation}
\frac {2 \kappa}{Q} > \left(\frac{5}{6} \right)^{1/2} \frac{m \sigma_{0} }{R}
\end{equation}

This last is an instability criterion which gives the values of $\sigma_{0}$
below which a certain m-symmetry pattern will begin to grow, for a given
rotation curve and value of Q. We can also obtain directly the ranges of
values $\Omega_{p}$ is expected to take, considering that this limit values 
will be given by the condition $\omega =0$ in equation (18). Since the pattern 
will develop only when the second term in the R.H.S. of eq.(18) dominates over the first,
we can start by ignoring the first term on the R.H.S. of this equation. For example, in a 
flat rotation curve region, with a circular velocity of 220 $km/s$ and a
$\sigma=30 km/s$, even with a $Q=3$, the first term above is only 1.7 \% of the second for $m=1$,
one has to go to $m=4$ for this first term to pass 20 \% of the second. We can hence estimate the 
limit values of $\Omega_{p}$ by neglecting this first term to give:

\begin{equation}
m \Delta \Omega_{p,lim} = \pm 2 \sqrt{2} \frac{\kappa}{Q}.
\end{equation}

Remembering that $\Delta \Omega = \Omega_{0} -\Omega_{p}$, we obtain:

\begin{equation}
\Omega_{p,lim} = \Omega_{0} \pm \left( \frac{2 \sqrt{2}}{Q} \right) \frac{\kappa}{m}.
\end{equation}

Since $Q$ is of the order of $2\sqrt{2}$, the above equation is a direct analytical explanation for the ranges of pattern speeds seen in
numerical experiments, independent of the swing amplification hypothesis.

Going back to equation(19), we see that the lower values of $m$ are easier to excite than the higher
ones, as often found in numerical simulations, providing an explanation
for observed galaxies with strong, high pitch angle arms, being preferentially
$m=2$ systems. Indeed, Martos et al. (2004) showed that the $m=4$ optical spiral pattern in the
Milky Way can form as a consistent hydrodynamical response of the gas to an underlying
$m=2$ pattern in the old stellar population, as traced by COBE-DIRBE K-band photometry.

Remembering that $\kappa/Q$ is really just a place-keeper for $(G \rho_{0})^{1/2}$, the instability 
criterion of eq.(19) in the original terms reads:

\begin{equation}
\frac{R}{m \sigma_{0}}>\left( \frac{5}{12 \pi}\right)^{1/2} \left( \frac{1}{G \rho_{0}}\right)^{1/2}
\end{equation}

In this terms, the instability criterion of eq.(22) is just a comparison between the 
local gravitational timescale of the disk, the free-fall timescale, and the inter-arm
diffusive crossing time. That is, the instability will not develop if diffusion is such that
the perturbation is blurred out before it can condense gravitationally. The lower the
number of arms, the easier it is to excite the pattern, as in high $m$ patterns, it is
easier for stars to diffuse from one arm to another, blurring and erasing the proposed
pattern. To conclude, we identify the dimensionless number

\begin{equation}
P_{m}=\left(\frac{5}{24} \right)^{1/2} \frac{m Q\sigma}{\kappa R} =\left(\frac{5}{12 \pi G \rho} \right)^{1/2}
\frac{m \sigma}{R}
\end{equation}

as the critical indicator for an n-armed quasi radial density wave to develop, with 
the instability threshold appearing at $P<1$.


The consequences of  eq.(23) for spiral arms
and bars will be explored in the following section, under two idealised regimes, 
flat rotation curve, and solid body rotation, for natural values of the parameters $m$ and $Q$.

\section{Consequences for Galactic Disks}
\label{sec:chemestry}

In this section we shall trace some of the consequences for the onset of the instability
presented, in the context of thin galactic discs. For simplicity, in this first treatment, we shall
consider only two distinct regimes for the rotation curve of a galaxy, a strictly flat rotation curve
regime, of relevance to the development of galactic arms, and a strictly solid body rotation curve
region, appropriate for the modeling of dynamics of bars. The above with the intention of 
developing as clear a physical understanding of the effect being introduced as possible.
A more detailed treatment of the problem using more realistic rotation curves and
detailed numerical galactic models, appears in section (4).

\subsection{Galactic Arms in Flat Rotation Curves}

We start this section with a consideration of the flat rotation curve limit of the relations derived
previously, as a suitable first approximation to the dynamics of a galaxy in the region over which 
galactic arms are seen. In a flat rotation curve regime, $\kappa = 2^{1/2} \Omega$, and therefore, eq.(19) becomes:

\begin{equation}
\frac{V_{r}}{\sigma_{0}} > \left( \frac{5}{3}\right)^{1/2}\frac{n Q}{4}.
\end{equation}

In real galactic systems, typically with $5 < V_{r}/\sigma_{0} < 8$, we see that in a flat rotation curve regime
the instability criterion will always be satisfied for $m=2$, for example, in a $Q=1$ or $Q=2$ disk. For $m=4$
it is still easy to satisfy the instability criterion for the onset of a strong perturbation,
for $Q<3$, but higher values of $m$ would be hard to accommodate. Indeed, in the detailed stability analysis 
of simple power-law disks in idealised rotation curves of Evans \& Read (1998), 
it is shown explicitly that lower $m$ modes are always the most unstable ones. Actually, from the radial
dependence of $P_{m}$, it can be seen that higher $m$ patterns will be easier to excite as one moves further
out in a galactic disk, in consistency with the bifurcations often seen in galactic spiral arms, patters
which tend to be characterised by higher order symmetries as one moves to larger radii.

It is interesting to note that
since Q scales linearly with $\sigma$, $P_{m}$ which scales with $Q \sigma$, is a quadratic function of $\sigma$.
This last means that for small values of $\sigma$, $P_{m}<Q$, with the opposite
being true for large values of $\sigma$. By setting $P_{m}=Q$ we obtain

\begin{equation}
\frac{4 V_{r}}{m \sigma} =\left(\frac{5}{3} \right)^{1/2},
\end{equation} 

for low values of m this condition will result in a value of $\sigma$ larger than 
that present in normal galactic disks, and hence we see that typically galaxies will
lie in the region $P_{m}<Q$. This is interesting, as perhaps the identification of
values of $Q \sim 3$ as necessary to stabilise galactic disks reported for numerical
experiments, might correspond to cases of $Q$ stable, $P$ unstable disks, with quasi-radial
density waves arising, and then being wound up into tight spirals.




Remembering again that $(\kappa/Q)^{2}$ is a proxy for $G \rho_{0}$, we can
compare equation (18) to the
classical WKB equivalent relation, e.g. eq.(6.61) Binney \& Tremaine (2008). We see that the term 
including $\sigma_{0}$ in equation (18) appears in replacement of the standard $\kappa$
term. By considering a purely radial perturbation with a purely tangential force field and 
resulting matter flows, we have lost the contribution of differential rotation, in what is
essentially an analysis at constant radius, in favour
of a dynamical pressure term through the
stellar velocity dispersion, which does not appear in the classical criterion.

\subsection{Central Bars in Solid Body Rotation Curves}

In moving to central galactic bars we now go to a solid body rotation regime. The above consideration 
leads to the use of $m=2$ in what follows. In this regime, $\kappa=2 \Omega_{0}$, with $\Omega$ a constant, 
and the instability criterion of eq(19) is modified by the substitution of a 6 for the 3 in the 
denominator of the square root term in the right. In terms of $\Omega_{0}$ this reads:

\begin{equation}
\frac{R \Omega_{0}}{\sigma_{0}} > \left( \frac{5}{6}\right)^{1/2} \frac{Q}{2}
\end{equation}

This time we see that the criterion for the onset of a quasi-radial perturbation
is guaranteed to fail inwards of a certain radius, as the term on the left scales
linearly with radius. Perturbations in the above regime will hence be in the weak oscillatory
linear regime of eq.(18). The limit values for $\Omega_{p}$ for these weak pulsating bars can be found from
eq.(18) setting $\omega=0$, which under solid body rotation gives:

\begin{equation}
\frac{\Omega_{p}}{\Omega}= 1 \pm \left[\frac{5}{3}\left( \frac{\sigma_{0}}{R \Omega} \right)^{2} 
-2\left( \frac{2}{Q}  \right)^{2}  \right]^{1/2}.
\end{equation}

\begin{figure}
\includegraphics[angle=0,scale=0.7]{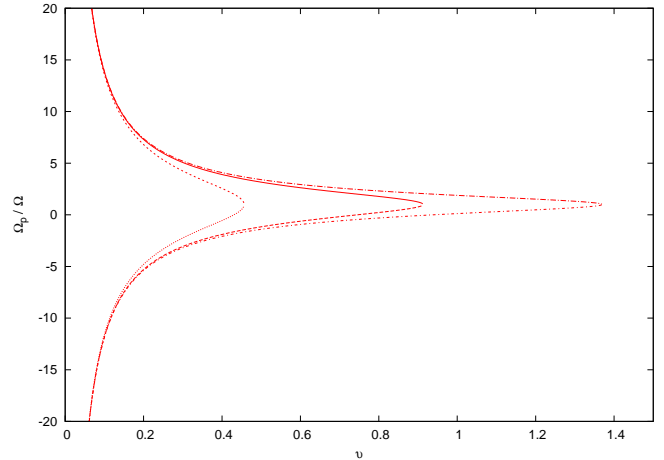}
\caption{Values of $\Omega_{p}/\Omega$ for stationary weak bars, as a function of the dimensionless
radial coordinate $\upsilon=R\Omega/\sigma_{0}$, for Q=1, inner curves, Q=2, middle curves and
Q=3, outer curves. The vertex of the curves defines the extent of the weak bars in question, always
at the corotation radius.}
\end{figure}

We find a slow and a fast solution, as a function of the radial coordinate, $\Omega_{0}$, $\sigma_{0}$ and $Q$.
Figure (1) now gives values of $\Omega_{p}/\Omega_{0}$, as a function of the dimensionless radial coordinate
$\upsilon=R\Omega_{0}/\sigma_{0}$, for 3 values of $Q$, 1, 2 and 3, inner, middle and outer
curves, respectively. Notice that for large values of $\upsilon$, relatively constant values of
$\Omega_{p}$ appear, for values of $Q>2$.
In going towards the centre however, the divergence towards $R=0$ becomes
evident, and large variations in $\Omega_{p}$ appear. This last feature might explain the appearance
of bars within bars, which exhibit increasingly rapid pattern rotation rates in going towards the 
central regions (e.g. Heller et al. 2007, Maciejewski \& Athanassoula 2008 and Shen \& Debattista 2008). 
We see that the lower branch of the curves, corresponding to the minus sign in
eq.(28), can have negative values over much of its extent. This corresponds to counter-rotating
bars, where the stars are not counter rotating themselves, but the density enhancement is.

Also, we see that replacing the inequality for an equality in eq.(27)
we can solve for $R_{M}$, the maximum extent of a weak bar of the type being described. This last
is given by the vertex of the curves in figure(1), and as can be seen from eq.(27), will
scale linearly with $Q$. From eq.(28) we see that these weak bars end at corotation, 
as already clear from eq.(21), the transition between the oscillatory and the unstable regimes
coincides with the corotation condition.
The interpretation provided here to eq.(22), highlighting the role of orbital diffusion, might offer a simple
explanation to the commonly found truncation of bars at corotation
in both orbital structure studies and N-body simulations, e.g. the unstable modes of Jalali \& Hunter (2005),
the simulated bars of O'Neill \& Dubinski (2003)
or the outer of the double bars of Maciejewski \& Athanassoula (2008) and Shen \& Debattista (2008).

In summary, weak bars can develop in the oscillatory regime of eq.(18),
and in the case of solid body rotation curves, present an almost constant non-winding $\Omega_{p}$
for a large range of radii, for values of $Q>2$, perhaps related to the 'pulsating bars' seen in some
numerical experiments, e.g. Shen \& Debattista (2007).

\section{Numerical Models and Simulations}

\begin{figure}[!t]
\includegraphics[angle=0,scale=0.64]{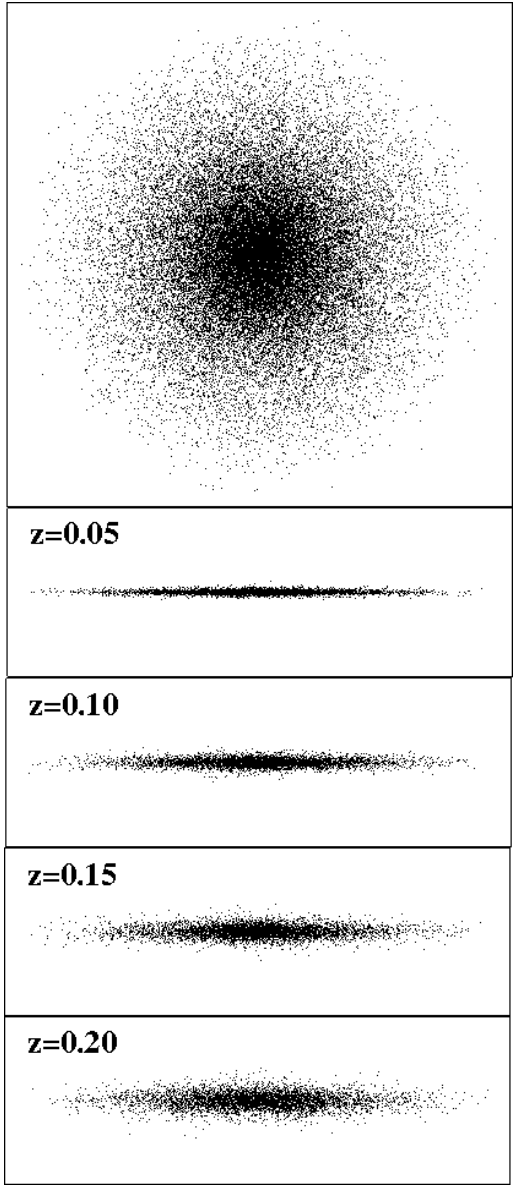}
\caption{Disk particles distribution at $T=0$. The upper panel is $12\times 12$ $LU$, and the other ones are $12\times 4$ $LU$.}
\label{fig:models}
\end{figure}

In order to check the predictions of the model presented in the more realistic cases of fully
self consistent 3D galactic models,
we have run a large grid of simulations of isolated, stable disk models. The models contain three self-consistent components, corresponding to a disk, 
a bulge and a halo. The construction  of the models and their following evolution were performed using the package Nemo (http://carma.astro.umd.edu/nemo/).

Setting up a stable three-dimensional disk for N-body experiments is a difficult task. Several strategies have been suggested including growing the disk 
mass distribution in a self-consistent halo/bulge model (Barnes 1988, Athanassoula, Puerari \& Bosma 1997), or treating the spherical components halo+bulge 
as a static background (Sellwood \& Merritt 1994, Quinn, Hernquist \& Fullagar 1993, Sellwood 2011). Some authors have also directly solved the Jeans 
equation for the complete disk, bulge and halo system to find the velocity dispersions (Hernquist 1993). The more important problem in the construction 
of self-consistent 3 component models is the fragility of a cool disk, generating grand design spiral structure and bars, as well as a series of other 
known instabilities (bending modes, disk warping, etc). 

\begin{figure}[!t]
\includegraphics[angle=0,scale=0.45]{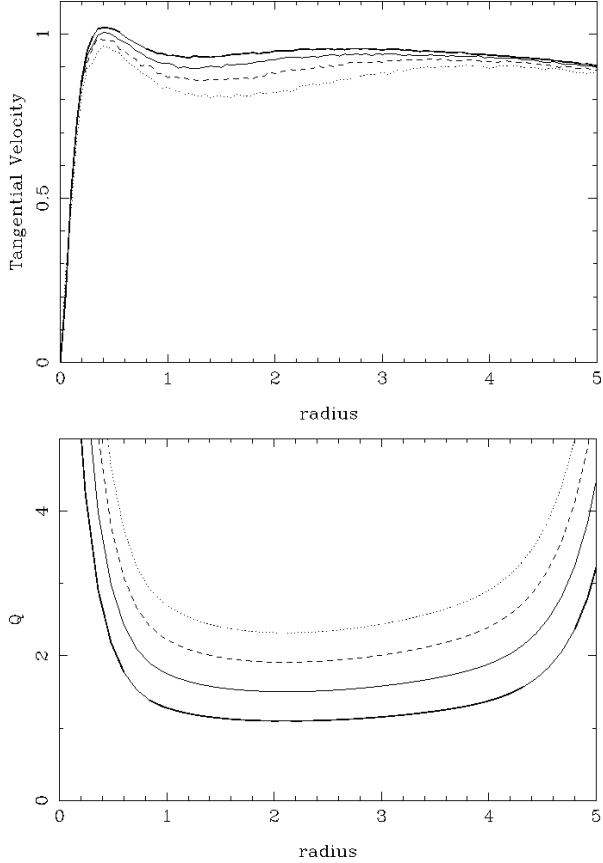}
\caption{Upper panel: tangential velocity curves at $T=0$. The thick solid line is for the model with $\sigma_{R,0}=0.27$. The solid, dashed and dotted lines are 
for $\sigma_{R,0}=0.37, 0.47,$ and 0.57, respectively. Bottom panel: Toomre's $Q$. The lines are as in the upper panel, i.e., the thick solid line represents the 
cooler model, and the dotted line represents the warmer one.}
\label{fig:velcurveq}
\end{figure}

Our models were constructed using the Nemo routine {\tt mkkd95}. The routine is based on the method described in Kuijken \& Dubinski (1995). Their strategy 
is the choice of analytic forms for the distribution functions (DF) of the three components. For the bulge, the DF is taking as a King model (King 1996). 
For the halo, the DF is a truncated Evans (1993) model for a flattened logarithmic potential. For the disk, the more difficult component to build 
up in the model, the authors used a vertical extension of the planar DF discussed by Shu (1969) and Kuijken \& Tremaine (1992). All the DF's are used to 
calculate the spatial density of each galaxy component, and the Poisson equation

\begin{equation}
 \nabla^2 \Psi(R,z)=4\pi G[\rho_{disk}(R, \theta, z)+\rho_{bulge}(R)+\rho_{halo}(R)]
\end{equation}

is solved by using spherical harmonic expansion by Prendergast \& Tomer (1970) with some modifications (see details in Kuijken \& Dubinski 1995).

\begin{figure}[!t]
\includegraphics[angle=0,width=8.0cm,height=10cm]{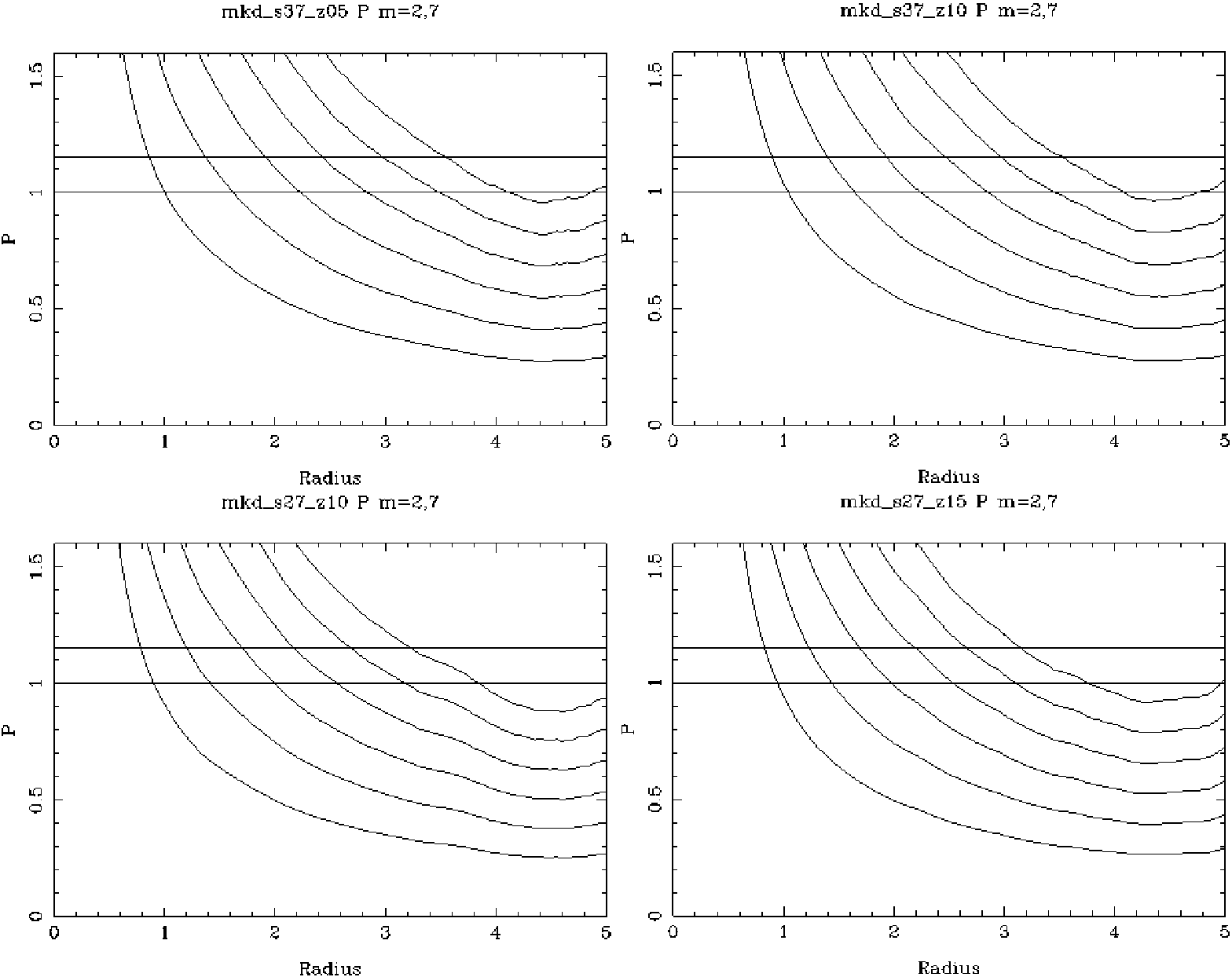}
\caption{The $P$ number from eq.(23) as a function of radius for 4 of our 16 simulations. Upper panels show the models with $\sigma_{R,0}=0.37$ ($z_0=0.05$, 
left; $z_0=0.1$, right), and bottom panels show colder models with $\sigma_{R,0}=0.27$, $z_0=0.1$, left; $z_0=0.15$, right. The plotted curves are the mean $P$ 
from $T=0$ to 200, 2Gyr. The lower curve in each panel is for $m=2$ and the upper one for $m=7$. The horizontal lines show limits for $P$ between 1.0 and 1.15. 
The upper and lower cuts of these limits and each $m$ curve constrain the radial region outwards of which we expect the power of that given $m$ to arise.}
\label{fig:ppp}
\end{figure}

\begin{figure}[!t]
\includegraphics[angle=0,scale=0.5]{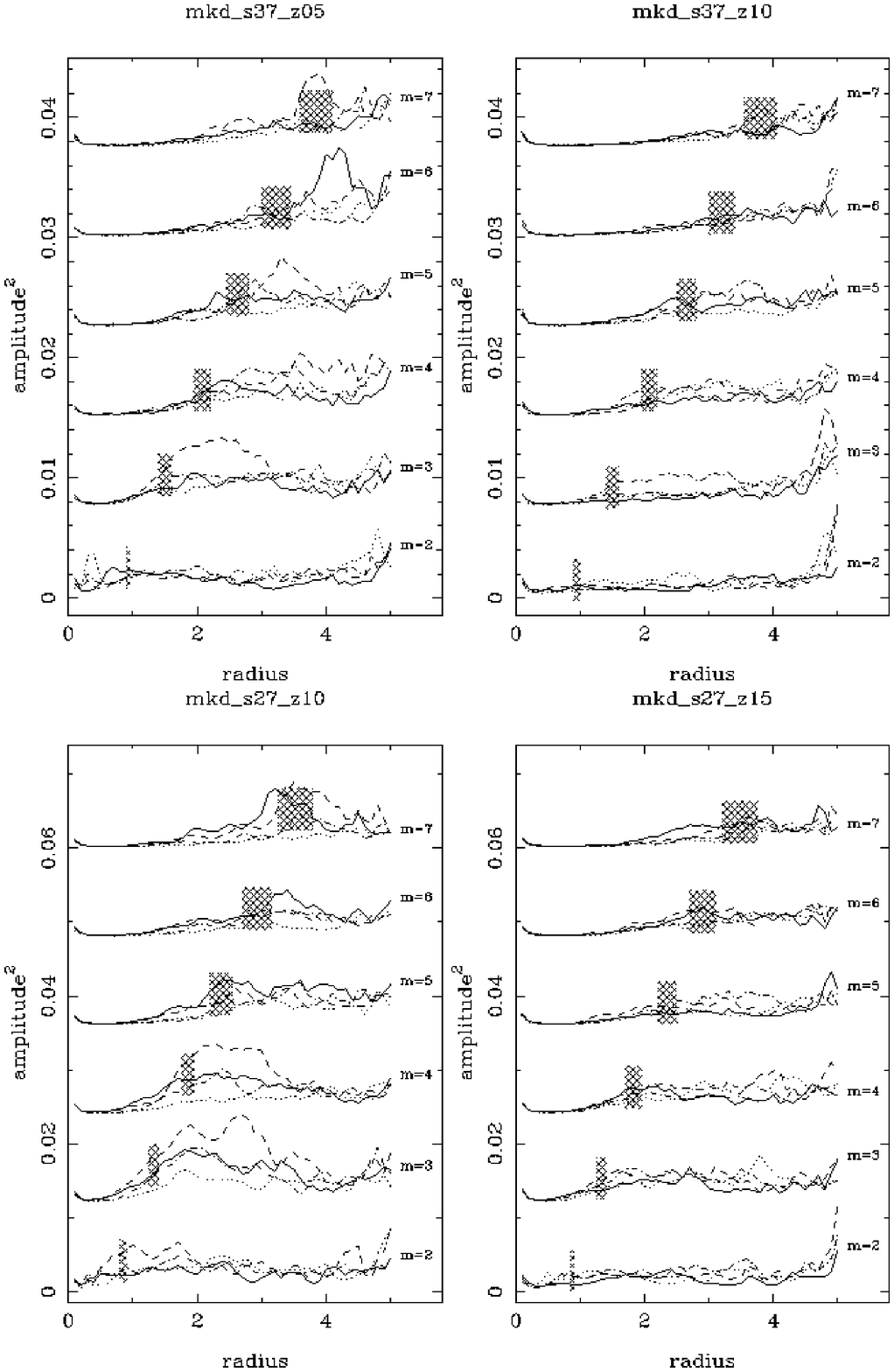}
\caption{Power of the $m$ components as a function of radius for the same simulations in figure (\ref{fig:ppp}). Different lines are the means for different time 
intervals: solid line, $0<T<50$, dashed $50<T<100$, dot-dashed $100<T<150$, and dotted, $150<T<200$. The hashed areas represent the radial range
outwards of which the power of a given component is expected to increase ($1<P<1.15$), for each $m$. For clarity, we added a constant to each 
subsequent $m$ curve above $m=2$.}
\label{fig:amplisquare}
\end{figure}

For the evolution of the models, we used the Nemo routine {\tt gyrfalcON}, which is a full-fledged N-body code using a force algorithm of complexity $ {\cal O}(N) $
(Dehen 2000, 2002), which is about 10 times faster than an optimally coded tree code. In the program, the gravitational constant is taken equal to 1. 
To match this value, an appropriate normalization has been chosen with length unit $LU=3$ kpc, and time unit $TU=10^7$ years. Using these values, the units 
of mass, velocity and volume density are $6\times 10^{10}$ M$_{\odot}$, 293 km s$^{-1}$, and 2.22 M$_{\odot}$ pc$^{-3}$, respectively, making the unit
of length 3 kpc. In this section, 
we plot all quantities in computational units. With {\tt gyrfalcON}, all the simulations were run using a tolerance parameter $\theta=0.5$, a softening 
parameter $\epsilon=0.05$  and a time step of $1/2^7$. These values ensure a total energy and angular momentum conservation of the order of $10^{-5}$ and 
$10^{-6}$, respectively.

Our models were chosen to be Milky Way A (MW-A) like-models from Kuijken \& Dubinski (1995), having masses of $M_D:M_B:M_H = 0.82:0.42:5.2$ and number of 
particles $N_D:N_B:N_H = 160,000:40,000:400,000$. This is twice the particle numbers which Kuijken \& Dubinski (1995) tested in their best model and showed 
to be in equilibrium. These authors show that disk surface density and velocities dispersion profiles hardly change with time when using the fully 
self-consistent model with 300,000 particles; only the disk scale length changes of the order of 15\%. All our models start with no vertex deviation,
and result in average values close to $\sigma_{\theta \theta} / \sigma_{RR} =0.75$ and  $\sigma_{zz} / \sigma_{RR} =0.6$ for the velocity ellipsoid.
The fiducial model of Kuijken \& Dubinski (1995) has a disk central radial 
velocity dispersion $\sigma_{R,0}=0.47$ and a vertical scale length $z_0=0.10$. We have created a grid of 16 models using $\sigma_{R,0}=0.27, 0.37, 0.47, 0.57$ 
and $z_0=0.05, 0.1, 0.15, 0.2$, and therefore covering the space parameter from cool thin disks, to warmer/thicker ones. Here we present results for 
colder/thinner models; in warmer/thicker models, the structures present much lower amplitude, and they are much more chaotic. Our fully self-consistent 
colder/thinner models support low amplification of several modes, unlike the unstable models, in which bars and m=2 grand design spiral patterns normally form.

In figure (\ref{fig:models}), we present a face-on view and edge-on snapshots for models with different vertical scale length $z_0$. The axisymmetrical aspect 
of the disk remains for all times, only changing about 15\% in vertical scale. In figure (\ref{fig:velcurveq}) we show the tangential velocity for models with 
different disk central radial velocity dispersion $\sigma_{R,0}$, as well as the initial Toomre's $Q$ profiles.

In figure (\ref{fig:ppp}) we show our profiles of $P$ for 4 models, 2 for $\sigma_{R,0}=0.37$ models (upper panels) and 2 for the colder 
model with $\sigma_{R,0}=0.27$ (bottom panels). For each $m$, the $P$ profile scales with $\sigma(R) \rho(R)^{-1/2}$, where $\sigma(R)$ is the disk radial velocity 
dispersion profile and $\rho(R)$, the disk volume density profile, c.f. eq.(23). The plotted curves are the azimuthally averaged time means of 
$P$ as a function of radius for the total time interval over which the models were evolved, from $T=0$ to $T=200=2 Gyr$. 
We have draw two horizontal limits for $P$, $P=1$ (lower straight line), and $P=1.15$ (upper straight line). These two limits are used 
to estimate the radial limit outwards of which a given $m$ component is expected to show a clear increase in its amplitude, as predicted
by the developments of the previous sections.

Now, to calculate the amplitude (or power) of each $m$ component as a function of radius, we divide the disk in 50 rings (from $R=0$ to $R=5$, $\Delta R=0.1$). 
For each ring, the 1D Fourier transform is calculated as:

\begin{figure*}[!t]
\includegraphics[angle=-90,scale=0.7]{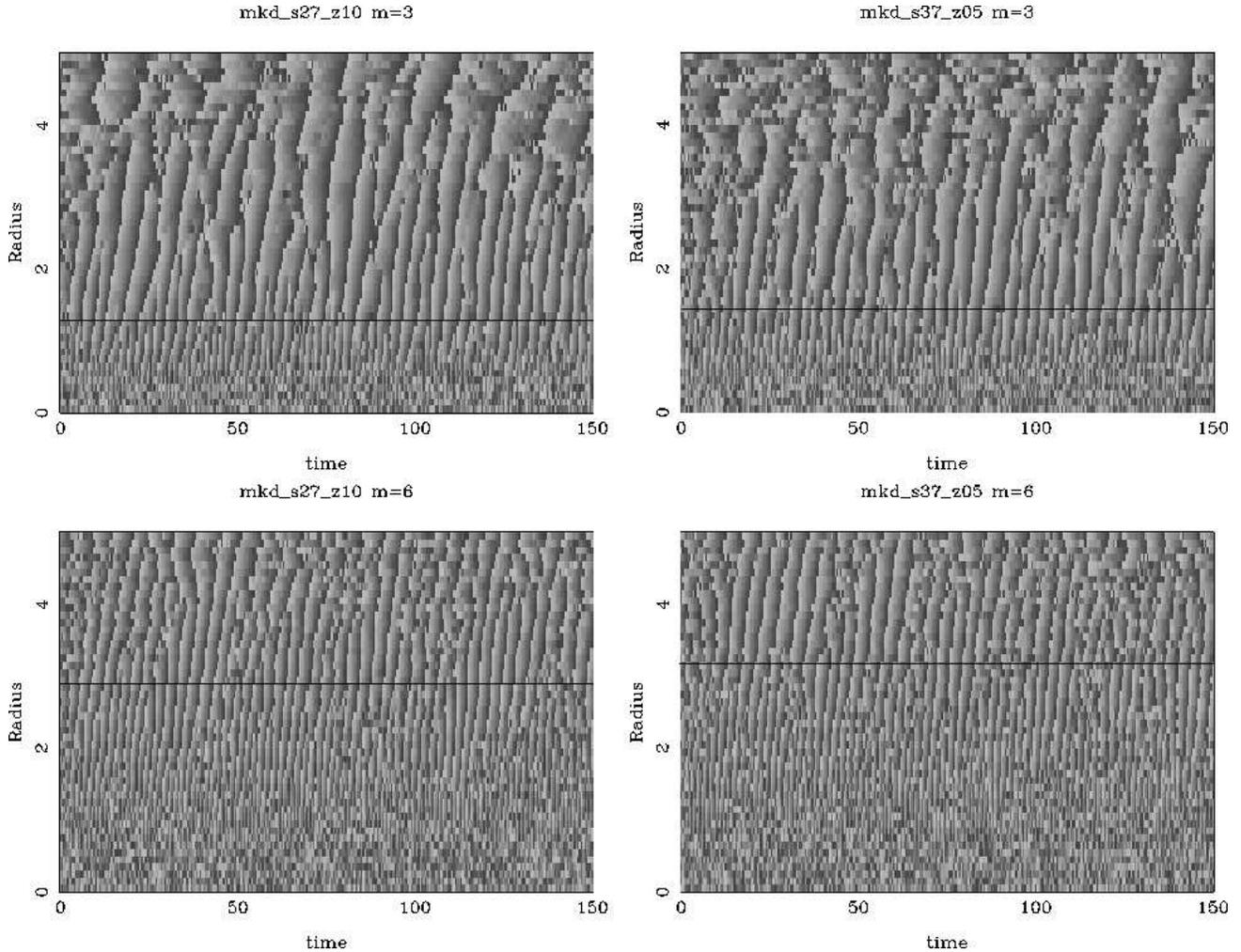}
\caption{The phase for the $m=3$ (upper panels) and $m=6$ (bottom panels) modes for two models appearing in Fig. \ref{fig:amplisquare}, as a function of time 
and radius (mkd\_s27\_z10, left; mkd\_s37\_z05, right). The phases range from 0 (black) to $2\pi/m$ (white). As all features are almost vertical, we see that 
the perturbations appearing are quite radial, far from tightly wound. The horizontal line for each $m$ of each model is the mean of the radial range show in 
figure (\ref{fig:amplisquare}). Note the clear change in the behavior of the phase on crossing the radial position marked, where we expect an increase of power
in terms of the theory presented.}
\label{fig:phase}
\end{figure*}

\begin{figure}[!t]
 \includegraphics[angle=0,width=8.5cm,height=16cm]{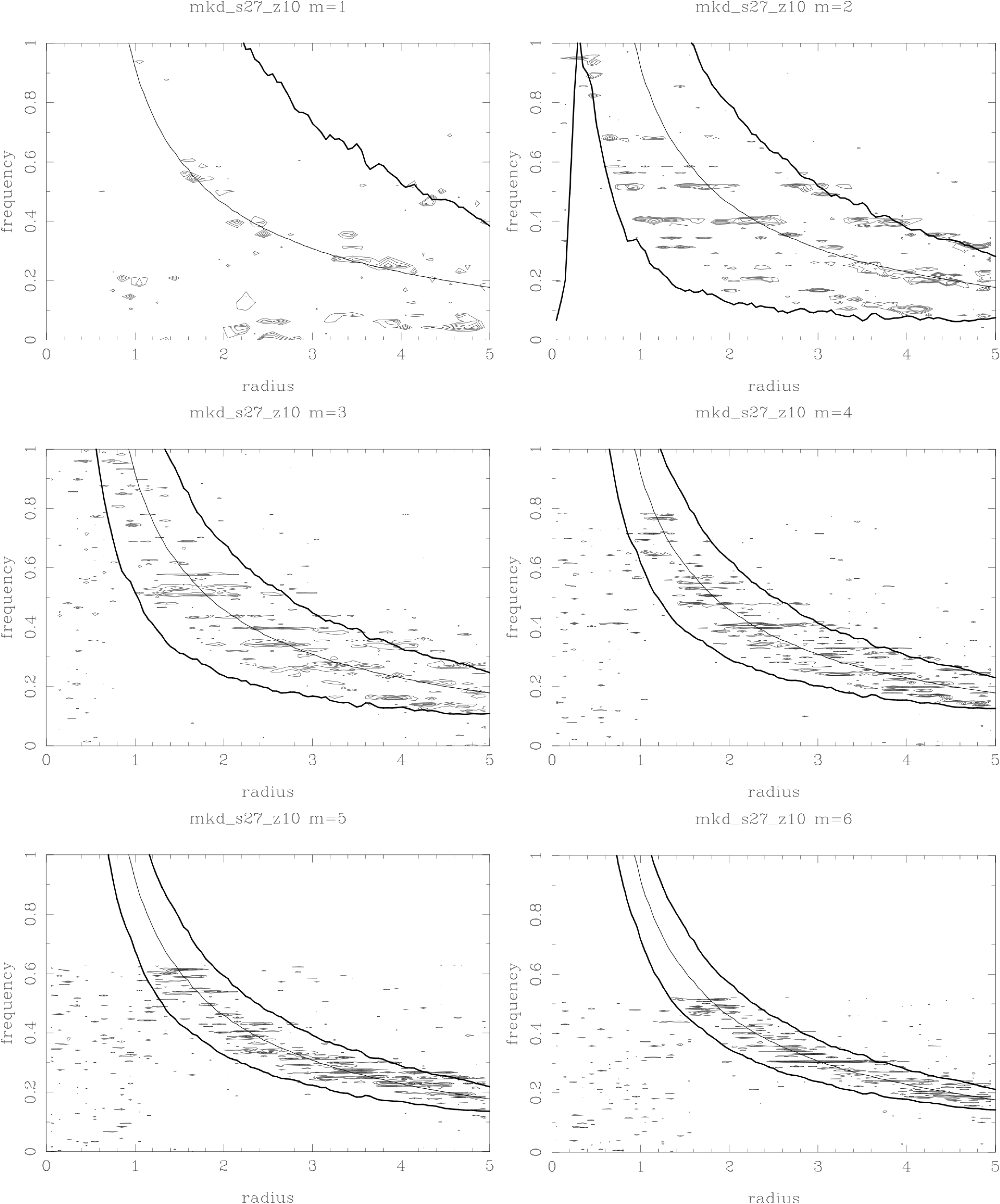}
\caption{The contour plots give the zones where power is present for the first 6 modes for one of the models, as a function of frequency and radius.
The three curves give $\Omega_{0}$ and $\Omega_{0} \pm \kappa/m$}
\label{fig:amplisquare}
\end{figure}

\begin{equation}
{\cal F}(m)={1\over N}\sum_{i=1}^N e^{-im\theta_i}
\end{equation}

where $N$ is the number of disk particles in the ring, $m$ is the $m$-armed component and $\theta$ the polar coordinate of each particle in the ring. The real 
and imaginary parts for each $m$ component are

\begin{equation}
I_c(m)=\sum_{i=1}^N cos(m\theta_i)\ \ \ \ \ \ \ \ \ \
I_s(m)=\sum_{i=1}^N sin(m\theta_i)
\end{equation}

and the power is then
 
\begin{equation}
Pow_m=I_c(m)^2+I_s(m)^2
\end{equation}

The phase $\Phi_m$ of each component is then:

\begin{equation}
\Phi_m=atan(I_s(m)/I_c(m))/m.
\end{equation}

We have calculated the power and phase of the first ten modes, as a function of radius and time. In figure (5), we show the power of the 
$m=2$ to 7 modes, as a function of radius. We present the mean value of the power for 4 time intervals. Superimposed on the curves, we have 
draw shaded areas representing the radial interval resulting from the analytic developments of section (2), as inferred through eq.(23) applied to figure (4).
There is a clear agreement between the radial positions from figure (\ref{fig:ppp}) and the radius at which we see an increase of power for the different $m$'s,
lending empirical support to the theoretical model presented in this study.

Next, we present some plots of phase as a function of radius and time in figure (\ref{fig:phase}). We present the phase of $m=3$ (upper panels) and $m=6$ 
(bottom panels) for the models mkd\_s27\_z10 (left) and mkd\_s37\_z05 (right). In these graphs, the phases range from 0 to $2\pi/m$. The 
amplitudes of the perturbations in our isolated models are quite low compared to those associated with bars and strong $m=2$ spiral arms in unstable 
models (Puerari, in preparation). 
Even with these low amplitudes, the phase shows a clear behaviour, quite ordered, representing quasi radial perturbations on the disk, as evident
from the almost vertical form of the features appearing. We also mark in figure (\ref{fig:phase}) the mean radial position of the corresponding
$P<1$ threshold of eq.(23), as taken from 
figure (\ref{fig:ppp}). At these radial positions we notice a change in phase behavior, from a more chaotic one (smaller radii) to more ordered 
features (larger radii). The radial position where this transition occurs agrees very well with the theoretical predictions, with this transition radii 
occurring first for the lower m modes, and only appearing at larger radii for higher m modes.

We end this section with figure (7), which shows the regions in $\Omega_{p}, R$ space where power appears for the different modes in one of the
simulations run. We see again a clear agreement with the theoretical expectations developed, with the amplitude of the modes constrained
to the region where $\Omega_{p}<\Omega_{0} \pm \kappa/m$. Comparing with the analytical predictions of eq. (21), and given the values of $Q$
of the simulated galaxies of $\sim 2$, we see again good a agreement. If one wanted to calibrate the details of the theory through
a comparison of eq.(21) with the results of figure(7), we see than a small numerical adjustment, e.g. in the first order scale height estimate
of eq.(14) of a factor of 2, would furnish an accurate accordance of the limit $\Omega_{p}$ predictions of eq.(21) and figure (7).

In summary, we see that even though in the theoretical developments of sections (2) and (3) we ignored the vertical structure of the galactic disk,
the radial motion of the stars, and the radical variations in surface density, epicycle frequency and velocity dispersion, the predictions for
the critical radii at which quasi-radial perturbations appear, the way in which these increase progressively for higher modes, and the
limit frequencies which these exhibit, are all in good accordance with the results of fully self-consistent 3D numerical models. 

\section{Conclusions}

We have developed a simple description of a quasi-radial density wave in a galactic disk,
assuming only tangential forces due to the perturbation, we reach a description of
the problem which is complementary to the classical tight-winding approximation. The main
features are the substitution of the differential rotation for the velocity
dispersion of the disk in the resulting dispersion relation. The resulting dispersion
relation allows a clear understanding of why lower $m$ modes are more unstable than higher
$m$ modes, and of why the higher $m$ values appear progressively at larger radii.

Weak bars in harmonic potential central regions appear in the oscillatory regime, an expression for $\Omega_{p}$ is 
found which shows a divergence towards $R=0$, perhaps helping to understand the phenomenon
of increasingly rapidly rotating bars within bars seen in numerical simulations. The extent of this
weak bars is seen to be naturally limited by the corotation condition. 

The expectations for quasi-radial spiral arms are tested in detail through extensive 
numerical simulations, which confirm to good accuracy the main results of the theory 
developed here.

\section*{acknowledgements}

This work was supported in part through CONACyT (45845E),
and DGAPA-UNAM (PAPIIT IN-114107 and IN103011) grants.

\end{document}